\documentstyle[preprint,aps,pre,psfig]{revtex}
\begin{document}

\newcommand{\be}{\begin{equation}}
\newcommand{\ee}{\end{equation}}
\newcommand{\BE}{\begin{eqnarray}}
\newcommand{\EE}{\end{eqnarray}}
\newcommand{\BEn}{\begin{eqnarray*}}
\newcommand{\EEn}{\end{eqnarray*}}
\newcommand{\barr}{\begin{array}} 
\newcommand{\earr}{\end{array}}
\newcommand{\bit}{\begin{itemize}}      
\newcommand{\eit}{\end{itemize}}
\newcommand{\bfl}{\begin{flusleft}}
\newcommand{\efl}{\end{flusleft}}
\newcommand{\bfr}{\begin{flushright}}
\newcommand{\bc}{\begin{center}}
\newcommand{\ec}{\end{center}}

\newcommand{\ben}{\begin{enumerate}}    
\newcommand{\een}{\end{enumerate}}

\newcommand{\cl}{\centerline}
\newcommand{\ul}{\underline}
\newcommand{\nl}{\newline}

\renewcommand{\theparagraph}{\Alph{paragraph}}

\draft

\title{Transition to Stochastic Synchronization in Spatially
Extended Systems}

\author{Lucia Baroni and Roberto Livi\footnote{Electronic Address: livi@fi.infn.it \,
URL : http://docs.de.unifi.it/$\sim$docs}} 

\address{Dipartimento di Fisica, Universit\'a di Firenze, 
        L.go E. Fermi, 5 - I-50125 Firenze, Italy\\
        Istituto Nazionale di Fisica della Materia, UdR Firenze,
        L.go E. Fermi, 3 - I-50125 Firenze, Italy}

\author{Alessandro Torcini\footnote{Electronic Address: torcini@ino.it\,
URL : http://www.ino.it/$\sim$torcini }\\[-2mm]}

\address{Dipartimento di Fisica, Universit\'a ``La Sapienza'',
        P.le A. Moro, 3 - I-00185 Roma, Italy\\
        Istituto Nazionale di Fisica della Materia, UdR Firenze,
        L.go E. Fermi, 3 - I-50125 Firenze, Italy}

\date{Draft version, \today}

\maketitle

\begin{abstract}
Spatially extended dynamical systems, namely coupled map lattices,
driven by additive spatio-temporal noise
are shown to exhibit stochastic synchronization.
In analogy with low-dymensional systems, synchronization can be 
achieved only if the maximum Lyapunov exponent becomes negative
for sufficiently large noise amplitude.
Moreover, noise can suppress also the non-linear mechanism
of information propagation, that may be present in 
the spatially extended system. A first example of phase
transition is observed when both the linear and the 
non-linear mechanisms of information production
disappear at the same critical value of the noise amplitude. 
The corresponding critical properties can be hardly 
identified numerically, but some general argument suggests that they
could be ascribed to the Kardar-Parisi-Zhang
universality class. Conversely, when the 
non-linear mechanism prevails on the linear one, another 
type of phase transition to stochastic synchronization occurs. 
This one is shown to belong to the universality class of directed 
percolation.
\end{abstract} 

\vspace{2mm}


\tightenlines

\begin{center}
{\bf Introduction}
\end{center}

The synchronization of dynamically coupled chaotic systems
is a new interesting phenomenon that has attracted many 
studies and efforts in the last years. The first models and
examples discussed in the literature concerned the case where
the dynamical coupling is performed by a deterministic chaotic
signal (either external or self-generated) acting on different 
trajectories of the same low-dimensional
chaotic system (e.g., see \cite{carroll}).
Under quite general conditions, e.g. large enough coupling strength
and sufficiently long evolution time, the two trajectories 
approach each other and, despite their chaotic nature, no more 
diverge.
Beyond the intrinsic conceptual interest of "chaotic synchronization",
the main motivations of these studies came from the possibility of 
identifying efficient
mechanisms for controlling or suppressing chaos in deterministic
signals, with promising perspectives of applications to the automatic
control of devices and to cryptography. 

On the other hand, applying a control mechanism to a deterministic
system by a chaotic signal looks 
artificial with respect to the more realistic scenario,
where the coupling could be produced by noise. It is worth stressing that
deterministic chaos cannot be assimilated to a random process, despite
in some cases they may yield equivalent results
\footnote{For instance, the Monte Carlo procedure applied
to an ergodic system yields the same equilibrium properties 
emerging from the Liouville measure associated with the dynamics.}.
Accordingly, it is quite natural to raise the question if synchronization
can be observed also when the chaotic coupling is substituted
with a stochastic signal. 
The problem of "stochastic synchronization" has been investigated by several 
authors \cite{pikov,lowd,pikov2,HF,Lai} considering two different initial
conditions of the same low-dimensional chaotic system (e.g. the logistic
map or the Lorenz system) and making them evolve by adding to both 
of them the same realization of noise. 
For sufficiently large values of the noise amplitude,
the stationary probability measure associated with the chaotic dynamics 
may be modified by the addition of noise favouring the
stabilization onto the same stochastic orbit.  
As argued by Pikovsky \cite{pikov},
this effect can be explained and quantitatively
predicted by observing that, at some value of the noise amplitude, the
Lyapunov exponent associated with the 
dynamics changes from positive to negative. 
A necessary condition for the occurrence of synchronization
is that the dynamical system has expanding and
contracting regions, so that noise may amplify the role of the
latter against the former. For instance, synchronization cannot
emerge from uniformly hyperbolic maps, like the Bernoulli shift.
In Section I we discuss ``stochastic synchronization'' in simple
maps, that we introduce as examples of different classes of models.
The new question that we want to address in this paper is 
the following: under which conditions stochastic synchronization
may occur in spatially extended dynamical systems coupled by the same 
realization of spatio-temporal noise? 
The models, that we shall study numerically,
are the diffusively coupled map lattice (CML) \cite{cml}
versions of those
introduced in Section I.
A comparison with the analysis therein worked out on 
stochastic synchronization in low-dimensional systems 
shows that the spatial structure makes this phenomenon
even richer.
Specifically, let us list hereafter the main results contained in 
this paper:

\begin{itemize}
\item{} Pikovsky's criterion \cite{pikov} can be generalized
to spatially extended systems : Stochastic synchronization occurs only if the 
maximum Lyapunov exponent $\Lambda$ \cite{benettin} is negative.
In fact, we find that for noise-coupled CML models $\Lambda$
becomes negative for values of the noise amplitude $\sigma$ larger than
a threshold value $\sigma_\Lambda$.

\item{} It is well known that in spatially extended systems 
information may propagate with a finite velocity $V$ \cite{kan2},
due to spatio-temporal instabilities. For noise-coupled CML
models, with smooth nonlinearities, the only relevant instability
mechanism is the linear one, so that $V$ vanishes
with $\Lambda$. This identifies a
synchronization transition (PT1) at a threshold value $\sigma_V \equiv
\sigma_{\Lambda}$, whose critical properties
seem to belong to the Kardar-Parisi-Zhang (KPZ) universality class 
of interface roughening transition \cite{kpz}.

\item{} We discuss also other models where
deterministic dynamics or spatiotemporal noise may induce
strong non-linear effects.
In these cases we find that $V$ vanishes at  a value of
the noise amplitude $\sigma_V > \sigma_\Lambda$.
A different kind of synchronization transition (PT2) is observed
at $\sigma_V$ and its critical properties are
found to belong to the universality class of 
directed percolation (DP)\cite{DK,grass_dp}.    

\item{} In the latter case we find also that 
the time needed to achieve a synchronized state grows 
exponentially with the system size $L$ for 
$\sigma_V > \sigma > \sigma_\Lambda$ : this dynamical regime is 
analogous 
to the "stable-chaotic" phase observed in \cite{CoupledMaps}.
Conversely, this time grows logarithmically with $L$ for
$\sigma > \sigma_V$ (see also \cite{noi}). 

\end{itemize}

In Section II we introduce the models and the suitable indicators
for describing the synchronization transitions induced by
spatio-temporal noise in CMLs.
The dynamical features of the different scenarios are analyzed 
in Section III. In Section IV, we study the critical properties 
associated with the transition to stochastic synchronization in 
spatially extended systems : the numerical analysis
indicates that this phenomenon is a genuine
nonequilibrium process, characterized by the scaling laws 
of KPZ and DP universality classes for PT1 and PT2,
respectively. We expect that a suitable coarse-graining or 
renormalization approach could provide a rigourous ground for these
results if one could map the noise-coupled CML dynamics
onto the stochastic PDE's that identify the above mentioned universality classes.
Unfortunately, we have not been able to work out any consistent 
derivation. Nonetheless,  numerics allows one for drawing some
conclusions about the problem at hand: they are summarized in
Section V, together with some perspectives for future work on this
subject.

\section{A Survey on Stochastic Synchronization in low-dimensional systems}

Before entering the main topic of this paper, i.e. stochastic
synchronization in spatially extended systems, it is worth
presenting concepts, tools and properties characterizing the
same kind of phenomenon in low-dimensional dynamical systems.
In this section we consider some simple models, that exemplify
the general features of this phenomenon.

The basic model equation is the stochastic map
\be
x^{t+1} =  f({x}^t)\,+\, \sigma\cdot \eta^t  
\label{stomap}
\ee
where the state variable $x^t$ is a real quantity depending on the discrete
time index $t$, 
$\sigma$ is the amplitude of the time dependent random variable $\eta^t$, 
uniformly distributed in the interval [-1,1],
and $f(x)$ is a map from ${\cal S} \subseteq  I\!\!R $ into the interval 
${\cal I} \subseteq I\!\!R $. 
Stochastic synchronization can be investigated by
considering two different initial conditions, $x^0$ and
$y^0$ of dynamics (\ref{stomap}) with the same realization of additive 
noise $\eta^t$.
More precisely, we assume that the corresponding trajectories, $x^t$ 
and $y^t$, generated by (\ref{map}) may synchronize if their 
distance
\be
d(t) \, =\, |x^t - y^t|
\label{distlow}
\ee
becomes smaller than a given threshold $\Delta $, usually assumed much smaller
than unit. This definition allows one to identify
two natural quantities associated with the synchronization of trajectories: 
the {\sl first passage time} $\tau_1(\Delta)$, i.e.
the first instant of time at which $d(t) \leq \Delta$ and
the {\sl synchronization time} $\tau_2(\Delta)$, i.e. the interval
of time during which $d(t)$ remains smaller than $\Delta$ \cite{nota1}.
We want to point out that, in general, both $\tau_1(\Delta)$ and 
$\tau_2(\Delta)$ depend on the initial conditions and on the realization 
of noise. Accordingly, their averages over both ensembles have to
be considered as the quantity of interest for our analysis.
For the sake of simplicity we shall indicate these averaged times
with the same notations $\tau_1(\Delta)$ and $\tau_2(\Delta)$.

In all the numerical simulations, for sufficiently
small value of $\Delta$ (typically $\Delta < 10^{-7}$), the
results do not depend on the choice of $\Delta$. Moreover, one can
assume that the trajectories have eventually reached 
the synchronized state if, after
$\tau_1(\Delta)$, $d(t)$ remains smaller than $\Delta$ for
arbitrarily large integration times. This amounts to say
that $\tau_2 (\Delta)$ goes to infinity with
the integration time. This heuristic definition of the synchronized state
can be replaced by a more quantitative criterion:
according to Pikovsky\cite{pikov}, for a given value of
$\sigma$ the synchronized state occurs if 
the Lyapunov exponent $\Lambda$ of dynamics (\ref{stomap})
is negative. This indicator is defined as follows: 
\be
\Lambda \, = \, \lim_{t\to\infty}{1\over{t}} 
ln \prod_{j=1}^t |{\xi^j\over{\xi^{j-1}}}|
\label{lyaplow}
\ee
where the dynamical variable $\xi^t$ obeys the "linearized" dynamics
\be
\xi^{t+1} =  {\partial x^{t+1}\over{\partial x^t}}\xi^t
\label{tanmap}
\ee
and the derivative is computed along the trajectory given
by (\ref{stomap})~.
We want to remark that $\Lambda$ is well defined 
for deterministic dynamics (e.g. the $\sigma = 0$ case
in (\ref{stomap})~).  In the framework of the linear stability analysis
a positive (negative) $\Lambda$
measures the average exponential expansion (contraction) rate of nearby 
trajectories.
For $\sigma \not= 0$ we are faced with stochastic trajectories and it is not 
{\sl a priori} obvious if $\Lambda$ is still a meaningful quantity
\cite{nota2}. 
On the other hand, eq.(\ref{tanmap})
does not depend explicitely on
$\eta^t$ and it is formally equivalent for both the noisy
and the noise-free cases. However, the presence of noise modifies the
evolution of the system w.r.t the noise-free case and, accordingly,
also the tangent space dynamics. We have verified numerically
that $\Lambda$ is a self-averaging asymptotic quantity also
for dynamics (\ref{stomap}) with $\sigma \ne 0$. 
It can be interpreted as the average
exponential expansion (contraction) rate
of infinitesimal perturbations of stochastic trajectories
generated by the evolution rule (\ref{stomap})~.
In particular, its value is found to depend on $\sigma$, but not
on the realization of noise.

Let us remark that it is quite simple to
argue why $\Lambda < 0$ implies $\tau_2 \to \infty$:
After some finite time $\tau_1 (\Delta)$, $d(t)$
has decreased below a small threshold $\Delta$ and
the trajectories can be viewed as a perturbation of each other. 
Linear stability implies that their distance will keep on decreasing 
exponentially with an average rate $\Lambda$, so that, within numerical
precision, they will converge rapidly onto the same trajectory.

As a first example, we consider the continuous map shown in Fig.~\ref{f01}:
\be
f(x) =
\left\{\begin{array}{ll}
 - c \tanh(b\,(1+x)) &\quad {\rm if} \enskip x < -1;\\
a\,x\,(1-|x|) &\quad {\rm if} \enskip  |x| <1;  \\
c \tanh(b\,(1-x)) &\quad {\rm if} \enskip x > 1.\\
\end{array}\right. 
\label{contmap}
\ee
where ${\cal S} \equiv $R and ${\cal I} = [-1,1]$. 
We choose the parameter values $a = 4$, and $c = 0.5$, so
that (\ref{contmap}) can be viewed as a sort of anti-symmetrized 
version of the logistic map at the Ulam point, taking values over the whole
real axis. It can be easily shown that for 
$\sigma = 0$ and independently of $b$, 
$\Lambda = \ln 2$, i.e. map (\ref{contmap}) is chaotic.
Notice that the noise term of amplitude $\sigma$
extends ${\cal I}$ to the interval $[-1-\sigma,1+\sigma]$. 
We have verified numerically that, for any value of $b$
and  for $\sigma$ larger than a threshold value $\sigma_\Lambda$, 
$\Lambda$ becomes negative and after some finite time $\tau_1$
a synchronized state is eventually achieved.
In particular, we find that $\sigma_\Lambda$
is strongly dependent on $b$: for instance, we have obtained $\sigma_\Lambda=1.2$
for $b=2$ and $\sigma_\Lambda=0.019$ for $b=1000$.
As recently found also by Lai and Zhou \cite{Lai} for a similar mapping, 
this result indicates that symmetric , i.e. zero-average, noise can yield 
stochastic synchronization.

It is worth mentioning that some time ago Herzel and Freund \cite{HF} 
conjectured 
that stochastic synchronization can be achieved only if noise has a non-zero
average. They were led to such a conclusion by studying stochastic
synchronization for the case of a map $f$
of the unit interval into itself, i.e. ${\cal S}\equiv {\cal I} =
[0,1]~$. In such a case the application of the stochastic evolution rule 
(\ref{stomap}) demands the adoption of some further recipe
for maintaining the state variable $x^t$ inside the unit interval,
when $\sigma \not= 0$.
For instance, one can choose the following {\sl reinjection}
rule $x^{t+1} \rightarrow x^{t+1} + 1$  
( $x^{t+1} \rightarrow x^{t+1} - 1$) 
if  $x^{t+1} < 0 $ ($x^{t+1} > 1$).
As discussed in \cite{HF}, any recipe of this kind yields an
effective state-dependent noise, that does not preserve the 
original symmetry of the stochastic process $\eta^t$, thus 
acquiring a non-zero average value.
The above described example and the results obtained in \cite{Lai}
disprove their conjecture. 

Nonetheless, an interesting observation
is contained in \cite{HF}: the stochastic
evolution rule (\ref{stomap}) for maps definite on a finite 
interval induces strong non-linear effects due to the 
discontinuities introduced into the dynamics by the state-dependent noise.
This strong non-linear character of the dynamics is irrelevant
for stochastic synchronization in low-dimensional systems;
conversely, it reveals a crucial 
property for discriminating between
different critical behaviours in high-dimensional systems,
as we shall discuss in Section IV.

It is also worth considering that even the presence 
of non-zero average noise does not necessarily guarantee
stochastic synchronization.
For instance, a counterexample is provided by
considering dynamics (\ref{stomap}) for the
logistic map at the Ulam point:
\be
f(x) = 4 x (1-x) 
\label{logistic}
\ee
whose noise-free dynamics is mixing.
The state-dependent noise modifies the probability measure
of $x^t$ in such a way to increase the weight
of the contracting regions of the map. On the other hand,
$\Lambda$ remains positive as well as $\tau_2$ remains finite
for any value of $\sigma \in [0,1]$,
despite they can be made so small and so large, respectively,
to produce an apparent
synchronization effect. As discussed in \cite{pikov2,HF},
misleading results can be obtained in this case due to the finite
computational precision of numerical simulations.

As a final example, we consider the map
\be
   f(x)  = \cases{ bx  & if $0 < x < 1/b$ \cr
                   a + c(x-1/b) & if $ 1/b < x < 1$}
   \label{noi} 
\ee
Its dynamics converges to a stable periodic attractor: for 
$b = 2.7$, $a=0.07 $ and $c=0.1$ this is a period-3 orbit with 
negative Lyapunov exponent, $\Lambda = ln(c b^2) \approx -0.316~$.
Numerical analysis shows that the Lyapunov exponent 
of the stochastic evolution rule (\ref{stomap}) remains negative for 
any value of $\sigma$ and, according to Pikovsky's criterion, 
a synchronized state is always achieved, as we have numerically checked.

In summary, we have presented in this section three different examples
of low-dimensional dynamical systems epitomizing possible scenarios
concerning the phenomenon of stochastic synchronization:
In the last example it occurs for any value of the noise 
amplitude $\sigma$, in the second one for no value of $\sigma$ 
and only above some threshold value, $\sigma_\Lambda$, in the first
example.

\section{Stochastic synchronization in CML models}

The generalization of the stochastic map (\ref{stomap}) to a CML model
\cite{cml}
with additive spatio-temporal noise can be 
defined by the following two-step evolution rule:
\begin{eqnarray}
& & {\tilde x}_i^t = (1 - \varepsilon) x_i^t + \frac{\varepsilon}{2} 
(x_{i-1}^t + x_{i+1}^t)  \cr
& & x_i^{t+1} =  f({\tilde x}_i^t)\,+\, \sigma\cdot \eta_i^t  
\label{map}
\end{eqnarray}
The real state variable $x_i^t$ depends now also on the discrete space 
index $i = 1,2,\cdots , L$: assuming unit lattice spacing, $L$ corresponds
to the lattice size.
The strength of the spatial coupling between nearest neighbour maps 
in the lattice is fixed by the parameter $\varepsilon$, that can take
values in the interval [0,1]. Note that this kind of coupling amounts
to a discrete version of a diffusive term. The application of 
the map $f$ mimicks the reaction term of reaction-diffusion PDE's,
so that CMLs are commonly assumed to represent a sort of 
discretized version of continuous PDE's.
At variance with standard CML models, dynamics (\ref{map})
contains also a stochastic term given by a set of identical, independent,
equally distributed (i.i.e.d.) random variables 
$\{ \eta_i \}$, whose amplitude is
determined by the parameter $\sigma$. In what follows these i.i.e.d.
random variables are assumed to be uniformly distributed in the interval 
[-1,1].
 
In full analogy with the low-dimensional case discussed in the
previous section, stochastic synchronization can be investigated by
considering two different initial conditions $\{x_i^0\}$ and
$\{y_i^0\}$ for dynamics (\ref{map}), coupled by the same realization of 
additive spatio-temporal noise.
More precisely, we assume that the corresponding trajectories, $\{x_i^t\}$ 
and $\{y_i^t\}$, generated by (\ref{map}) may synchronize if their distance
\be
d(t) \, =\, \frac{1}{N} \sum_{i=1}^N |x_i^t - y_i^t|
\ee
becomes smaller than a given threshold $\Delta $, usually assumed much smaller
than unit. 
Upon this definition one can straightforwardly extend to the CML case
indicators like the {\sl first passage time} $\tau_1(\Delta)$
and the {\sl synchronization time} $\tau_2(\Delta)$.
Also in this case we assume that the trajectories  synchronize if 
after the time $\tau_1(\Delta)$, $\tau_2(\Delta)$ is found to diverge
with the integration time.
In all the numerical simulations we have used the same value of
$\Delta$ as in the low-dimensional case. In fact, we have verified
that also for the high-dimensional case the results of numerical 
simulations are not affected by the choice of $\Delta$,
provided it is small enough, tipically $\Delta < 10^{-7}$.
Again, both $\tau_1(\Delta)$ and $\tau_2(\Delta)$ are quantities 
averaged over initial conditions and over realizations of noise.

In analogy with low-dimensional systems, we expect that the suitable 
dynamical indicator for identifying stochastic 
synchronization in dynamics (\ref{map}) is the maximum Lyapunov exponent, 
$\Lambda$. In particular the phenomenon is expected to occur 
only for values of the noise amplitude $\sigma$ such that 
$\Lambda$ is negative.
For what concerns the interpretation of this indicator for the stochastic
CML dynamics (\ref{map}), the same kind of remarks and conclusions
discussed in Section I for the low dimensional case still hold.
All the numerical estimates of $\Lambda$ have been performed by applying 
the standard algorithm outlined in \cite{benettin}. 
 
Another relevant indicator, strictly related to the spatial 
structure of CML dynamics, is the average propagation velocity of finite 
amplitude perturbations \cite{kan2}
\be
V = \lim_{t\to\infty} \lim_{L\to\infty} 
\frac{\langle N(t)\rangle}{2 t}
\label{eq:velf}
\ee
where 
\be
N(t) = \sum_{i=1}^L h_i^t \quad {\rm with} \quad
h_i^t = \
\left\{\begin{array}{ll}
1 &\quad {\rm if} \enskip |x_i^t-y_i^t| >  0 ;  \\
0 &\quad {\rm otherwise} \\
\end{array}\right. 
\label{front}
\ee
is the number of non-synchronized or ``infected'' sites at time $t$.
Here $\{ x_i^t \}$ and 
$\{ y_i^t \}$ represent the trajectories generated by dynamics
(\ref{map}) starting from two initial conditions that differ of finite
amounts $ \delta_i \sim {\cal O}(1)$ only inside a space region of 
size $S$:
\be
y_i^0 =
\left\{\begin{array}{ll}
x_i^0 + \delta_i &\quad {\rm if} \enskip |L/2-i| \leq S/2 ;  \\
x_i^0 &\quad {\rm otherwise} \\
\end{array}\right. 
\label{finam}
\ee
The average $<\cdot>$ in (\ref{eq:velf}) 
is performed over different initial conditions and noise realizations.
The indicator $V$ measures the rate of information propagation : we
want to point out that $V$ can take finite values even for non-chaotic 
evolution, i.e. for $\Lambda < 0$. 
For instance, this scenario has been observed for
CMLs made of discontinuous maps \cite{CoupledMaps}.
In this case infinitesimal perturbations
are unable to be amplified, while finite amplitude
perturbations, induced by the discontinuity, can propagate
thanks to the spatial coupling.
Therefore in spatially extended systems 
the information flow is absent only when $V$ vanishes. 
 
As a first example let us consider 
the stochastic CML dynamics (\ref{map}), equipped with map (\ref{contmap})
for $b=2.0$ (see Fig.\ref{f01}). In this case, numerical simulations indicate
that the stochastic synchronization of trajectories occurs for any value of the
diffusive coupling $\varepsilon$ and for sufficiently large values of
$\sigma$, above which $\Lambda$ becomes negative.
For instance, with $\varepsilon = 1/3$ we find that synchronization is 
obtained for $\sigma > \sigma_\Lambda = 2.4768$. 
This exemplifies the validity of Pikovsky's criterion also for spatially 
extended systems. 

Let us provide more details of the dynamics below and above the 
threshold value $\sigma_\Lambda$.
For $\sigma < \sigma_\Lambda$ we find that
$\tau_1(\Delta)$ diverges exponentially with the system's size
$L$, i.e. $\tau_1 \sim \exp{(L/\xi)}$. The length scale factor
$\xi$ is found to be independent of $L$ and proportional
to the inverse decay rate of the space correlation
function of $x_i^t$. Accordingly, $L/\xi$ is an estimate
of the number of effective independent
degrees of freedom. In this dynamical regime the probability
$P(\xi,\Delta)$ that two trajectories get closer than
a distance $\Delta$ is proportional to the combined probability
that each one of the $L/\xi$ degrees of freedom 
get closer than $\Delta$, i.e. $P(\xi,\Delta) \propto \Delta^{L/\xi}$.
One can reasonably assume $\tau_1^{-1} \propto P(\xi,\Delta)$;
this rough argument explains why $\tau_1 \sim exp(L/\xi)$.
Note that even if $d(t)$ eventually becomes smaller
than $\Delta$ synchronization is rapidly lost because of
the linear instability mechanism due to positive $\Lambda$
and $\tau_2$ is always finite.

Conversely, a logarithmic dependence of $\tau_1(\Delta)$ on
$L$ is found for $\sigma > \sigma_\Lambda$ (see Fig. \ref{f02} (b)).
Numerical simulations show that in this case the number of 
regions made of a few synchronized sites increases as time flows.
Moreover, once a synchronized region is formed, it
grows linearly in time, until all regions merge and 
synchronization sets in over the whole lattice. 
One can introduce an argument  accounting for 
the logarithmic dependence $\tau_1 \propto \ln(L)$.
An effective rate equation for the number of synchronized sites,
$n(t)$, can be constructed by assigning a
probability $p$ for the formation of new synchronized sites and
a rate $\gamma$ for the linear increase of synchronized regions:
\be
\label{eqrate}
\frac{\partial n}{\partial t} =
\gamma + p(L-n) \quad ,
\ee
with $0 \le n(t) \le L$. This equation can be solved with the initial
condition $n(0)=0~$, so that an estimate of
$\tau_1$ is obtained by imposing the condition $n(\tau_1) = L$:
\be
\tau_1 = \frac{1}{p} \ln \left[ \frac{pL}{\gamma} +1
\right] \quad .
\label{loga}
\ee
Note that a logarithmic dependence on $L$ 
is consistent with the condition $\frac{pL}{\gamma} >> 1$,
that can be satisfied for sufficiently large values of $L$.
By estimating the parameters $p$ and $\gamma$ directly from
numerics we have verified that they fit reasonably well the simple
phenomenological eq. (\ref{loga}).

The indicator $V$ does not provide any additional information about 
this kind of synchronization transition.
Actually, $V$ is found to be positive for $\sigma < \sigma_\Lambda$, 
while it vanishes at $\sigma = \sigma_\Lambda$. This implies that
in this model the linear mechanism of information production 
associated with a positive $\Lambda$ rules also the 
propagation of finite amplitude perturbations \cite{grass}.
In summary, for $\sigma > \sigma_\Lambda$ $\Lambda$ is negative 
and $V = 0$~, so that after the trajectories have reached
$\tau_1$ it seems that no information production mechanism
can be responsible of the resurgence of $d(t)$ above the threshold
$\Delta$. This is confirmed by numerical simulations, although
in principle one could not exclude that the occurrence of a 
temporarily positive exponential expansion rate at some lattice site might 
produce a local amplification of $d(t)$. 
In this respect, we want to remark
that $\Lambda$ is a global indicator, i.e. the 
exponential expansion rate between nearby orbits 
averaged in time and over the whole phase space. Accordingly
a negative value of $\Lambda$ is fully compatible with 
the above mentioned local, instantaneous event.
This point, that reveals important for the understanding of the
dynamical mechanisms underlying the synchronization
transition, will be analyzed more carefully in the following section.

Different scenarios are obtained by considering dynamics
(\ref{map}) with $f$ given by the logistic map (\ref{logistic}).
For $\varepsilon = 1/3$ and large enough values of $L$, one
recovers features very similar to the case of a single logistic map,
where the synchronization transition is absent.
In fact, for any value of $\sigma$, $\Lambda$ and $V$ are positive, 
while $\tau_1$ and $\tau_2$ remain finite.
Moreover $\tau_1$ is found to diverge exponentially with $L$
with a parameter dependent rate.

A different situation occurs for $\varepsilon = 2/3$, where
$\Lambda$ vanishes at $\sigma_\Lambda = 0.27$, while $V$ remains positive
up to $\sigma_V \approx 0.4$ (see Fig.~\ref{f1}).
According to Pikovsky's criterion one expects that synchronization 
occurs above $\sigma_\Lambda$.This is indeed the case, although, for 
$\sigma_\Lambda < \sigma < \sigma_V $~,
$\tau_1(\Delta)$ is still found to increase exponentially with $L$,
while for $\sigma > \sigma_V$~, $\tau_1(\Delta)$
grows logarithmically with $L$ (see Fig.~\ref{f2}).
Despite the strong analogy between this transition at $\sigma_V$
and the
one occurring in the first example at $\sigma_\Lambda$, 
we want to remark that there is a crucial difference
between them:
below threshold $\tau_2$ diverges in the former case, while it
is finite in the latter.

This indicates the existence of a new kind of
synchronization transition at $\sigma_V$ essentially ruled by $V$.
Let us point out that, at variance with the first example,
for $\sigma_\Lambda < \sigma < \sigma_V$
the non-linear mechanism of information propagation 
is enough for maintaining the exponential dependence of $\tau_1$ on $L$.
On the other hand, after
$\tau_1$ two trajectories get very close to each other and the 
negative $\Lambda$ stabilizes them onto the same 
stochastic trajectory.

Let us stress that the above described scenarios are not peculiar 
of the specific choice of the parameter values 
$\varepsilon = 1/3$ and $\varepsilon = 2/3$.
We have checked that their main features 
are robust w.r.t. small but finite variations of $\varepsilon$,
although a systematic investigation of the parameter space
would demand too huge computational times.

Upon these examples one can conclude that 
in a CML of very large but finite size $L$
stochastic synchronization of two trajectories within an accessible
time span occurs when $V$ vanishes.

Moreover, a very similar situation is obtained when considering dynamics 
(\ref{map}) for the model of period-3 stable maps (\ref{noi}) introduced
in \cite{CoupledMaps}.  
In this CML model $\Lambda$ is found to be negative, independently
of the value of the diffusive coupling parameter $\varepsilon$ and,
accordinlgy, the dynamics eventually approaches a periodic attractor.
On the other hand, the model exhibits a transition between a frozen
disordered phase with $V = 0$ to a chaotic phase with $V > 0$ at
$\varepsilon_c \approx 0.6$. The peculiar feature of this transition
is that these two phases are separated by a small {\sl fuzzy} region 
centered around $\varepsilon_c$, where both positive and null values 
of $V$ can be observed up to available numerical resolution. 

The addition of noise to this CML dynamics according to 
eq. (\ref{map}) has interesting consequences: $\Lambda$ is kept negative
independently of the noise amplitude $\sigma$, while a small amplitude 
noise destabilizes the frozen disordered phase: $V$ becomes positive
even for $\varepsilon < \varepsilon_c$, so that the {\sl fuzzy} transition  
disappears. This notwithstanding, by increasing $\sigma$ up to a
critical value $\sigma_V(\varepsilon)$~, $V$ is found to drop again 
to zero not only below, but also above $\varepsilon_c$. For instance, one 
has $\sigma_V \approx 0.16$ for $\varepsilon = 0.58$ and 
$\sigma_V\approx 0.18$ for $\varepsilon = 0.62$.
In both cases we recover the 
same kind of mechanisms characterizing the synchronization transition 
discussed for coupled logistic maps with $\varepsilon = 2/3$.

We want to point out that a finite value of $V$, when $\Lambda$ is
negative, is usually reported as a typical signature of a strong 
non-linear effect \cite{grass}. For instance, it has been shown 
\cite{CoupledMaps} that  the discontinuity of map (\ref{noi})
yields such an effect already for the noise-free
CML dynamics. Even if the discontinuity of map (\ref{noi})
is removed by interpolating
the expanding and contracting regions with a sufficiently
steep segment the effect is maintained \cite{poltor}.

The {\sl reinjection} mechanism introduced by additive noise
in maps of the interval [0,1] into itself produces similar discontinuities
in the dynamics. As we have shown this non-linear 
effect is sufficient for giving 
raise to dynamical phases with $\Lambda < 0$ and $V > 0$ in the noisy
dynamics (\ref{map}).

All these observations suggest to investigate 
if a similar scenario can be obtained
for map (\ref{contmap}) in dynamics (\ref{map}), by introducing an
almost-discontinuity in the map, since in this case the {\sl reinjection} 
mechanism is not present.
This is easily obtained by taking a sufficiently large value of
the parameter $b$, e.g. $b \sim {\cal O}(10^3)$. Numerical simuations
show that, still for $\varepsilon = 1/3$, it exists a range 
of $\sigma$-values for which $V >0$ and $\Lambda < 0$.  
In Fig.\ref{f5} we show the dependence of $\tau_1$ on $L$ in two regions
of the parameter space where $\Lambda$ is negative,
while $V$ is either positive or null.  If $V >0$
an exponential increases of $\tau_1$ with $L$ is
again observed, while a logarithmic dependence characterizes the
dynamical phase with $V=0$. 

Accordingly, this second kind of transition is not just an artifact
due to discontinuities introduced by the rejection rule,
but a mere consequence of sufficiently strong
non-linear effects, that may be produced also in the absence of
any rejection mechanism and also if the noise distribution
maintains its symmetry.                                                   

\section{Critical properties of the synchronization transition in CMLs}

In the previous Section 
we have described two different kinds of phase transitions for
stochastic synchronization where the noise amplitude $\sigma$
plays the role of the control parameter.
When both $\Lambda$ and $V$ vanish at $\sigma = \sigma_\Lambda$ 
the transition is from a non-synchronous dynamical phase to a 
synchronous one: we have denoted it with PT1.
In the other case, that we indicate with PT2, $\Lambda$ passes
from positive to negative values while $V$ remains positive
up to $\sigma = \sigma_V$, where we have observed a transition between 
different synchronous dynamical phases, characterized by an exponential 
and by a logarithmic dependence on the system size of the first passage 
time $\tau_1$ .

In both cases we have found that the correlation length
$\xi$, defined in Sect. II, diverges at the transition point
: this is an indication in favour of a continuous phase transition. 
Nonetheless, $1/\xi$ is found to vanish
less than linearly when $\sigma \to \sigma_\Lambda^-$ in PT1,
while it vanishes linearly in PT2 when
$\sigma \to \sigma_V^-$ (see Figs. \ref{f3} (a),(b)). 

Moreover, close to the critical point
the time averages of the mean distance between trajectories,
$z^t= \frac{1}{L} \sum_{i=1}^L |x_i^t-y_i^t|$,
and of the topological distance, 
$\rho(t) \equiv \frac{1}{L} \sum_{i=1}^L h_i^t $  
($ h_i^t = 1$ if $|z_i^t| > \Delta$, otherwise $h_1^t = 0$~)
exhibits a continuous dependence on $\sigma$
(for the sake of space in Fig.\ref{f8} 
we show these quantities only for the case of coupled logistic maps).

We want to remark that our definition of the synchronized state implies that
it exists a stable stochastic orbit that prevents
the trajectories of the dynamical system to flow apart
from each other below some very small, but finite 
threshold $\Delta$ so that $\tau_2 = \infty$. 
In both cases numerical simulations show that 
this is what happens above the critical point,
where $\Lambda < 0$ and $V = 0$. 
If at some lattice site $i$ a fluctuation makes the local Lyapunov 
multiplier positive (i.e.,  $\ln |f^\prime({\tilde x}_i^t)| > 0$, 
where $f^\prime$ indicates the first derivative of the map
w.r.t. its argument),
giving rise locally to the exponential divergence of nearby 
orbits, the process is rapidly reabsorbed due to 
the lack of any mechanism of information propagation.
In this sense the synchronized state should be equivalent to
the absorbing state, typical of Directed Percolation (DP) processes 
\cite{DK,grass_dp}.

Below the critical points the role of fluctuations determines the difference
between PT1 and PT2.
The inspection of the space-time evolution of dynamics
(\ref{map}), using the symbolic representation of the state variables
$h_i^t$, is quite helpful for visualizing such a difference.
For what concerns PT1, when $\sigma \to \sigma_\Lambda^- $ one observes that 
non-synchronized clusters propagate as time flows (since $V$ is positive) 
and, even if some
of them may eventually die, in the meanwhile new ones have started to 
propagate, emerging also from previously synchronized regions.  
Since also $\Lambda > 0$ any local fluctuation of $d(t)$ produced
by a positive multiplier has a finite probability to be amplified
and eventually propagated through the lattice.

On the contrary, sufficiently close to PT2,
for $\sigma \to \sigma_V^-$, non-synchronized clusters never emerge from
already synchronized regions and any connected non-synchronous 
cluster eventually dies at $\tau_1(L,\sigma)$ in a lattice
of finite size $L$ (a situation very similar to what is observed
in the active phase of DP as a finite size effect).
Even if, in principle, the
non-linear mechanism of information propagation is active, 
this suggests that, inside an already synchronized region,
any local fluctuation of the Lyapunov multiplier towards positive
values never persits long enough to activate
the non-linear process of information propagation.

Upon these numerical observations and exploiting the analogies with
other critical phenomena, we are led to conjecture not only
the existence of an absorbing, i.e. synchronized, state but also
that PT2, at variance with PT1, should belong to the
universality class of DP.
This can be confirmed only by direct measurements of the 
critical exponents associated with the synchronization transitions.
It is worth stressing that numerical estimates 
have been performed by approaching the critical points from below,
i.e. for $V \to 0^+$. 

Here we report the analysis of PT2 in the case of coupled logistic maps
with $\varepsilon = 2/3$. 
As usual a reliable measurement of any critical
exponent demands a very accurate estimate of the critical point,
i.e. $\sigma_V$, in this case. For this reason we have performed
careful simulations for evaluating the dependence on the system
size $L$ of $\tau_1$, that corresponds to the absorption 
time in the DP language.
At the critical point $\sigma=\sigma_V$, this time should
diverge as 
\be
\tau_1 (L,\sigma_V) \sim L^z 
\label{zeta}
\ee
where $z = \nu_\|/\nu_\bot$ is the dynamical exponent \cite{DK,grass_dp}.
The quantity $\tau_1(L,\sigma)$ is reported in Fig.\ref{f9} as a function
of $L$ in a log-log scale for different values of $\sigma$. 
The best scaling behaviour is 
obtained for $\sigma_V = 0.4018$~, where one has $z = 1.55 \pm 0.05$.
This result agrees quite well 
with the most accurate numerical estimates of the DP value 
$z = 1.5807$ \cite{jensen}.
Relying upon this result, we have also measured the
critical exponent associated with the temporal decay of the density 
$\rho(t)$ of active sites, i.e. those sites where $|x_i^t - y_i^t|
> \Delta$~. The DP transition exhibits at the critical
point the following scaling law for the topological distance:
$\rho(t) \sim t^{-\delta}$, with $\delta = \beta /\nu_\| = 0.1595$
\cite{jensen}. 
The density $\rho=\rho(t)$ is shown in Fig. \ref{f10} for three
different system sizes, namely $L=500, 1500$ and $2000$. 
As expected the scaling region increases with $L$ and 
for $L = 2000$ an optimal fitting in the interval
$ 3.2 \le \log(t) \le 4.8$ provides the estimate for the
critical exponent $\delta = 0.159 \pm 0.002$. 
This confirms that PT2 belongs to the universality class of DP.

For what concerns PT1 we have considered coupled maps of the type
(\ref{contmap}) for $b=2.0$ and $\varepsilon=1/3$.
The best scaling for the $\tau_1$ as a function of $L$ 
(accordingly to equation (\ref{zeta})) is observed for a noise
amplitude $\sigma_\Lambda = 2.5015$. For $\sigma = \sigma_\Lambda$
we have obtained the following estimates for the critical
exponents : $z \approx 1.01 - 1.04$ and $\delta \approx 0.35$.
Such values certainly do not correspond either to DP, 
or to any known universality class for percolation or growth processes. 
This seems analogous to what has been pointed out by
Grassberger \cite{grass_dp} for systems exhibiting
{\it incomplete deaths}: when the asymptotic state
is not a truly absorbing one the critical properties of DP
cannot be recovered.

Since in this situation an absorbing state seems not
to exist, the dynamical exponent $z$ can be
better estimated by
measuring the number of ``infected'' sites of the
chain $N(t)$ defined in (\ref{front}).
At the critical point $\sigma=\sigma_\Lambda$
the following scaling law is expected to hold
\be
N (t) \sim t^{1/z} \quad .
\label{damage}
\ee
The data from numerical analysis are reported in Fig. \ref{f11}. 
For short times ($ t < 500$) we   obtain the inverse of the
the dynamical exponent $1/z \sim 0.47 \pm 0.05$,
a value consistent with the one expected for
the Edwards-Wilkinson universality class 
$z_{EW} = 2.0$ \cite{barabasi}.
For longer times we observe a crossover
to a lower $z$-value that is consistent
with the one expected for the 1d KPZ universality class
(namely $z_{KPZ} = 3/2$). 
We think that these results could be interpreted in
terms of the conjecture
reported in \cite{grass3}. In that paper it has been
suggested that, for generic synchronization transition
of coupled spatio-temporal chaotic systems
with continuous state variables,
the appropriate universality class should
be the one of the KPZ model with a non-linear
growth limiting term \cite{grass_dp,munoz}.
This idea originates from the observation that, 
close to the synchronization transition, 
it is possible
to describe the dynamics of small perturbations in
terms of a reaction diffusion model with multiplicative noise
\cite{grass_dp}. Finally, this model can be mapped via a Hopf-Cole
transformation into an equation that corresponds to a KPZ
model with a non-linear term that prevents the surface from
growing indefinitely \cite{pikok}. The critical scaling laws
for this kind of models
have been reported in \cite{munoz}: the dynamical exponent $z$
is found to coincide with the KPZ-one, while the other
exponents are found to be different from the standard
KPZ ones.
Unfortunately a full numerical comparison of PT2
with the model discussed in \cite{munoz} is prevented by two major
technical problems: the extreme difficulty to estimate
with sufficient precision the critical value $\sigma_\Lambda$
and the strong finite size effects.

\section{Concluding remarks}

We have studied the problem of stochastic synchronization
induced by additive spatio-temporal noise in CML models.
In analogy with the low-dimensional case, synchronization of trajectories
is observed if the maximum Lyapunov exponent becomes negative at
a critical value of the noise amplitude.
We have also identified two different critical behaviours
associated with the synchronization transition.
One of them belongs to the universality class of Directed Percolation.
While the other one is not clearly identified, but indications suggest
that it could belong to the universality class
of the KPZ model with a non-linear growth limiting term.

In particular, the DP-like phase transition describes the crossover 
between an "active" and an "absorbing" phase.
The former is characterized by an exponential dependence
on the system size $L$ of the time needed for achieving the
synchronized state, while the latter exhibits a logarithmic
dependence. This scenario is reminescent of the phenomenology
associated with stable-chaos in CMLs \cite{CoupledMaps}, where
the dynamics approaches a periodic attractor, rather than
a "synchronized" stochastic trajectory.

The two kinds of synchronization transition here reported
are quite general for extended dynamical systems, since 
analogous behaviours have been recently observed for two coupled
CMLs without any external noise \cite{pikov3}.

Moreover, as far as the control of chaos is
concerned when the erratic behaviours present in the 
extended system are due solely to non-linear mechanisms 
(as it happens when the maximal Lyapunov is negative,
but $V$ is still positive) the control schemes based
on linear analysis \cite{ogy} should fail and new 
``non-linear'' methods have to be introduced.
We believe that the appropriate indicator to employ 
in this context is the Finite Size Lyapunov Exponent \cite{fsle},
because it is able to capture
infinitesimal as well as finite amplitude perturbation growth.

We also expect that the observed phenomenology is not
peculiar of CMLs and the present study can 
apply to a wider class of spatially extended dynamical
systems, like coupled oscillator or PDE's.

\acknowledgments{
We want to thank all the components of the D.O.C.S 
research group of Firenze for stimulating discussions
and interactions, and in particular F. Bagnoli and A. Politi.
We express our gratitude also to V. Ahlers, M.A. Mu{\~n}oz, 
A. Pikovsky and P. Grassberger 
for helpful discussions (P.G. also for providing us the efficient 
random number generator that we have employed in numerical simulations).
A.T. acknoweledges the contribution of T. Caterina, H. Katharina,
H. Daniel amd T. Sara for providing him a realistic realization of
a chaotic and noisy enviroment.
Part of this work was performed at the Institute of Scientific Interchange
in Torino, during the workshop on `` Complexity and Chaos '', June 1998
and October 1999. We acknowledge CINECA in Bologna and INFM for providing 
us access to the parallel computer CRAY T3E under the  
grant `` Iniziativa Calcolo Parallelo''.
}

%
%
%
%

\begin{figure}[h]
\cl{
\psfig{figure=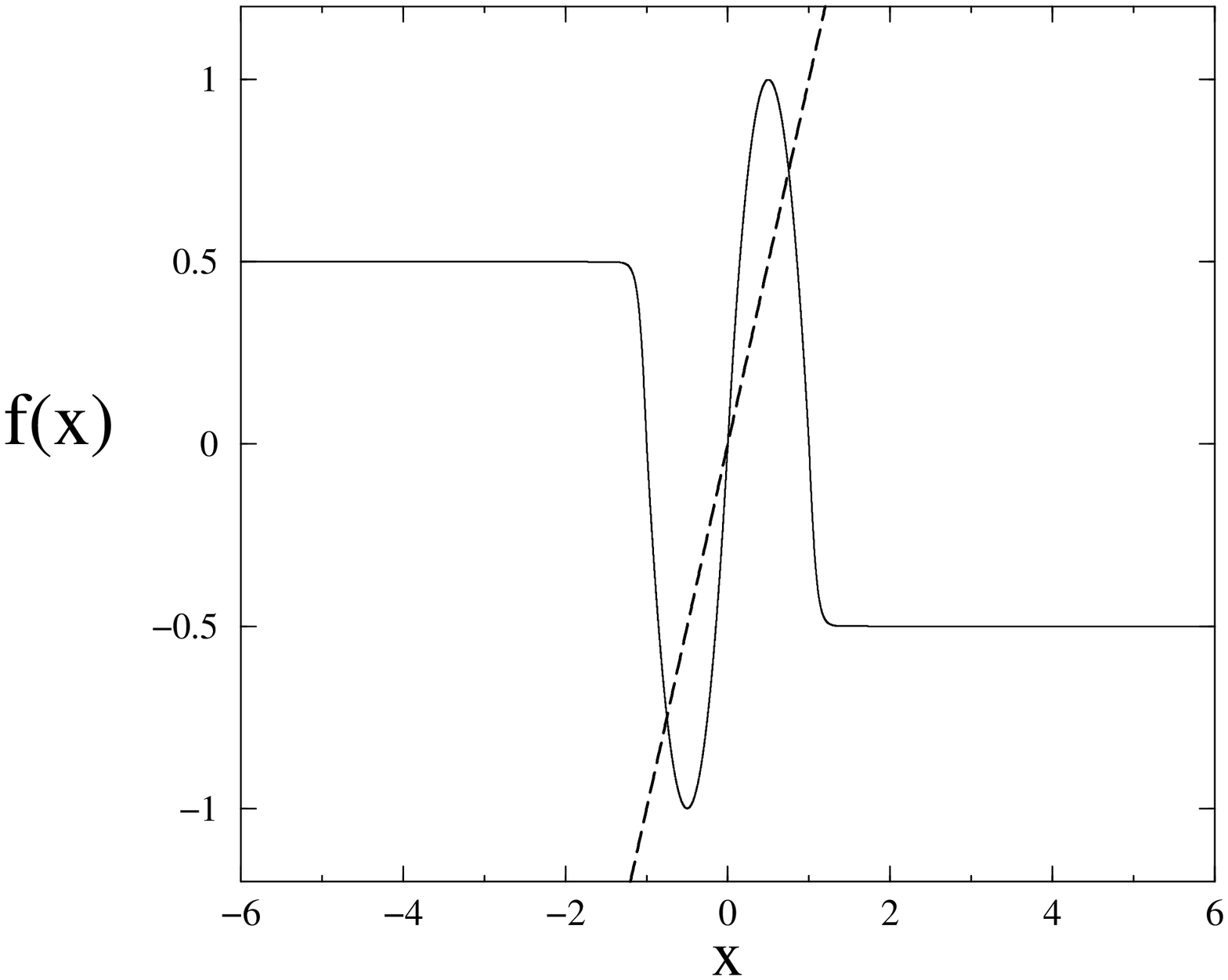,angle=-90,height=7truecm,width=7truecm}
}
\vskip 0.5 cm
\caption{
The map (\protect\ref{contmap}) for $a=4$, $b=2$ and
$c=0.5$.
}
\label{f01}
\end{figure}

\begin{figure}[h]
\cl{
\psfig{figure=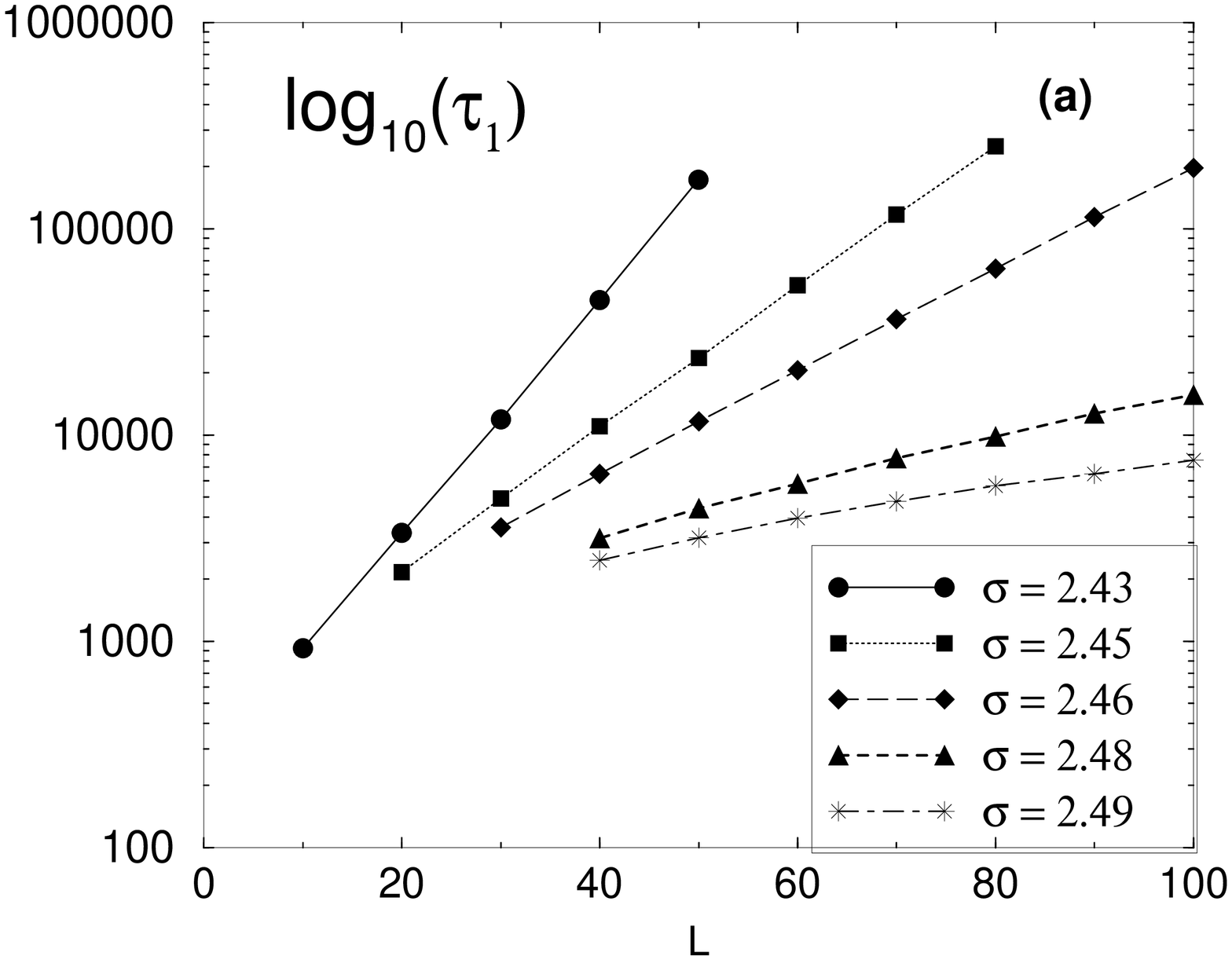,angle=-90,width=7truecm,height=7truecm}
\psfig{figure=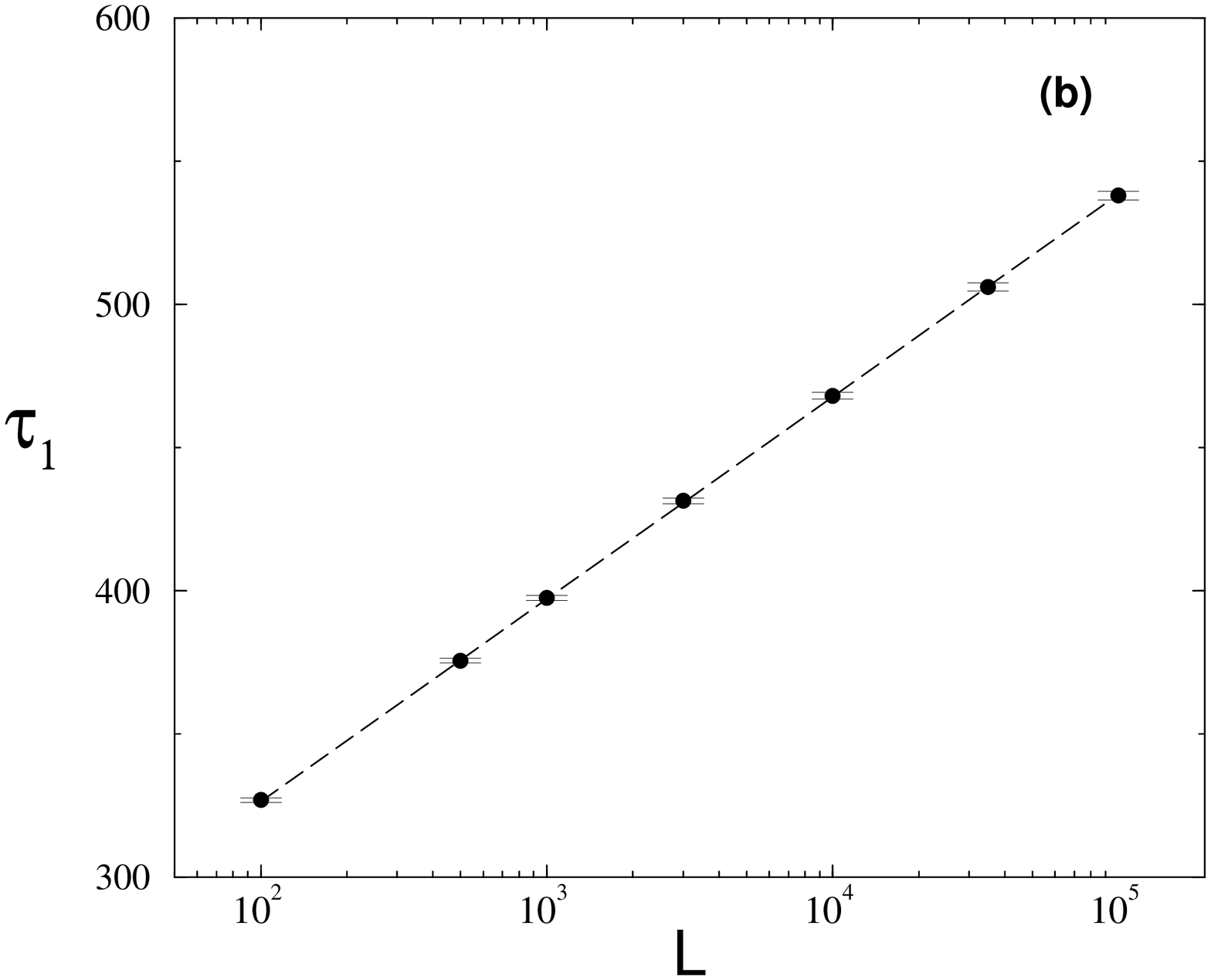,angle=-90,width=7truecm,height=7truecm}
}
\caption{$\tau_1$ versus $L$ for coupled maps (\protect\ref{contmap})
for $a=4$, $b=2$, $c=0.5$, with $\varepsilon = 1/3$:
a) exponential scaling observed for $2.49 \le \sigma \le 2.43$;
b) logarithmic scaling for $\sigma$ = 2.6 (filled circles).
The values of $\tau_1$ have been computed
with $\Delta = 10^{-8} - 10^{-12}$ and averaged over
$1000$ - $5000$ initial conditions.}
\label{f02}
\end{figure}

\begin{figure}[h]
\cl{
\psfig{figure=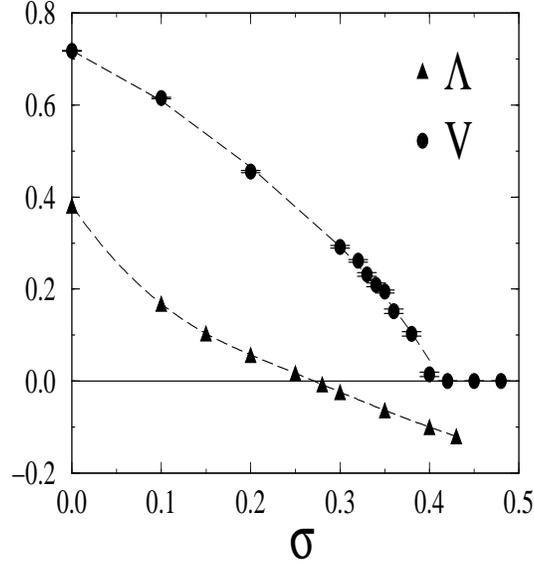,angle=0,width=7truecm,height=7truecm}
}
\vskip 0.7 cm
\caption{Behaviour of the propagation velocity $V$ (circles) and
of the maximum Lyapunov exponent $\Lambda$ (triangles) versus  
noise amplitude $\sigma$ for coupled logistic maps with
$\varepsilon = 2/3$. Both quantities have been computed for $L=1024$,
averaging over $10^3$ initial conditions each one followed for 
${\cal O}(10^5)$ time steps.
The dashed lines are a guide for the eyes.}
\label{f1}
\end{figure}

\begin{figure}[h]
\cl{
\psfig{figure=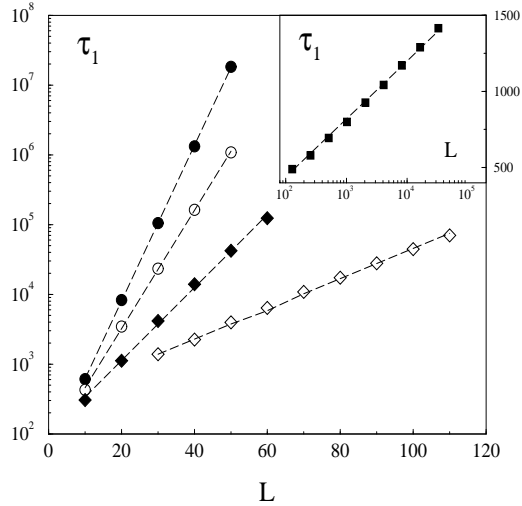,angle=270,width=7truecm,height=7truecm}
}
\vskip 0.7 cm
\caption{Exponential scaling of $\tau_1$ versus $L$ (reported 
in a lin-log scale)
for coupled logistic maps with $\varepsilon = 2/3$ at
different values of the noise amplitude: $\sigma$ = 0.3 (filled circles),
0.32 (empty circles), 0.35 (filled diamonds), 0.38 (empty diamonds).
The inset shows the logarithmic scaling of $\tau_1$ versus $L$ 
(in log-lin scale)
for $\sigma = 0.45$~. The values of $\tau_1$ have been computed
with $\Delta = 10^{-10}$ and averaged over
$10^3$ initial conditions.}
\label{f2}
\end{figure}

\begin{figure}[h]
\cl{
\psfig{figure=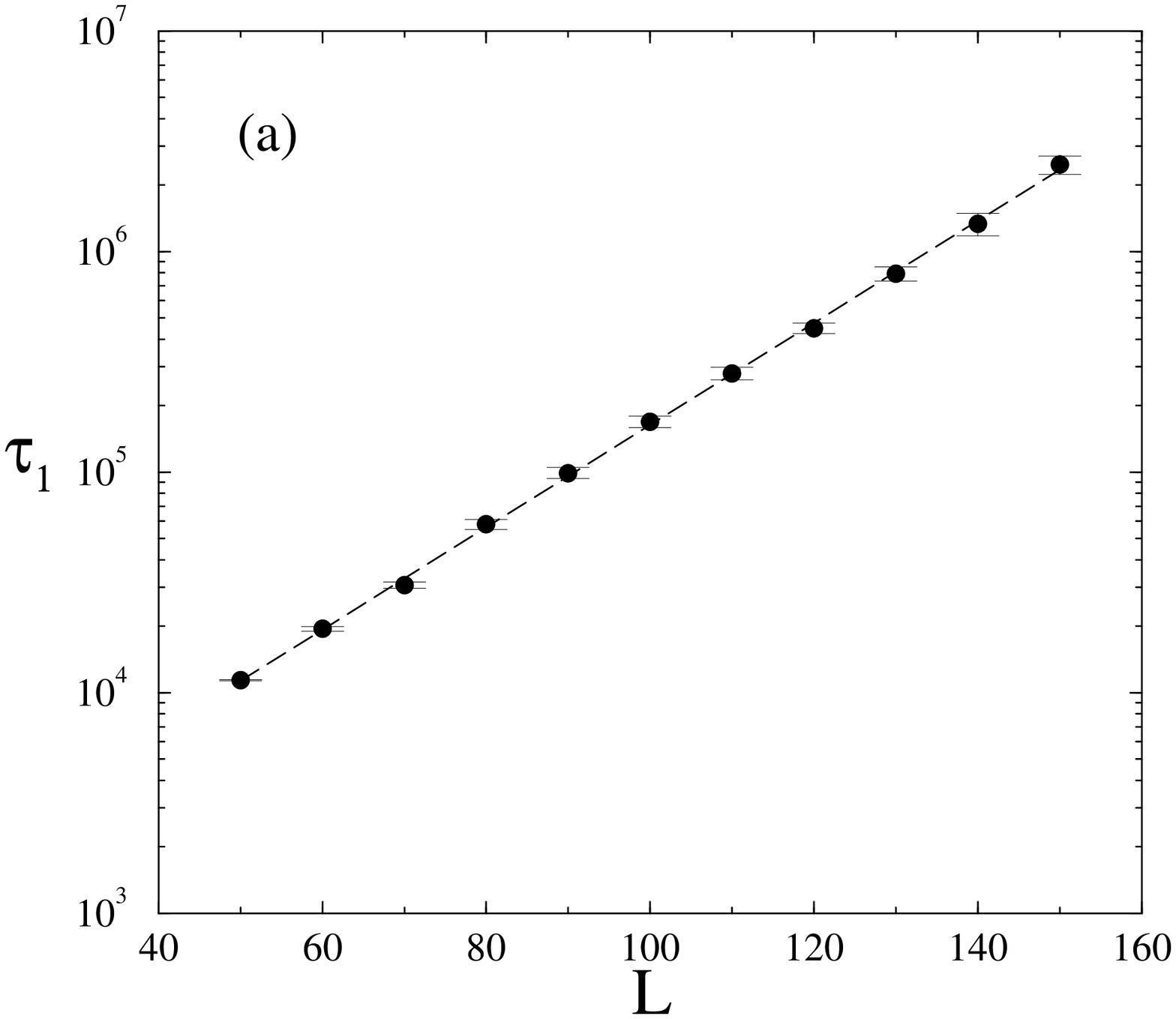,angle=-90,width=7truecm,height=7truecm}
\psfig{figure=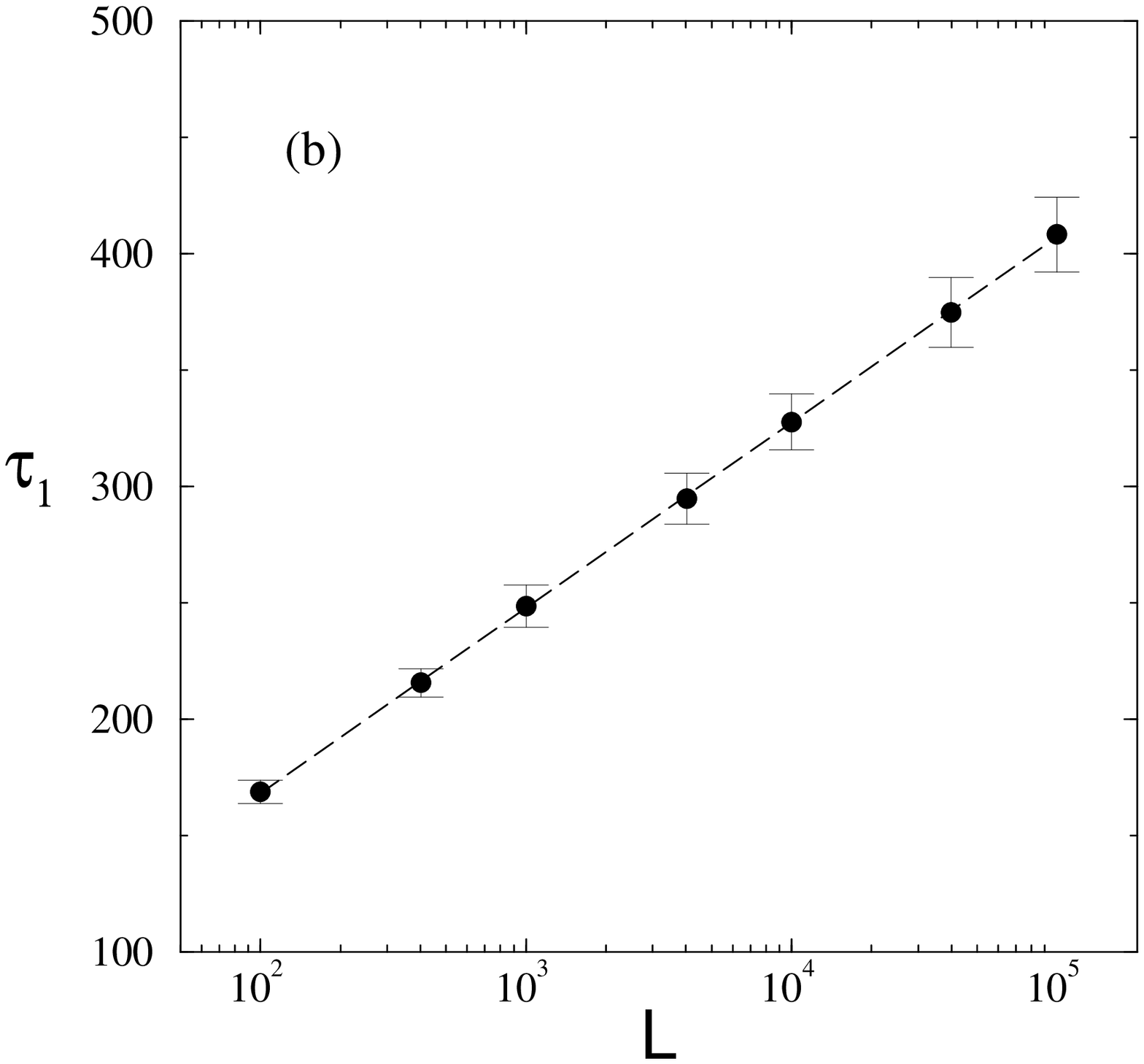,angle=-90,width=7truecm,height=7truecm}
}
\caption{(a) Exponential scaling of $\tau_1$ versus $L$ 
(reported in lin-log scale) for
coupled maps (\protect\ref{contmap}) for $a=4$, $b=1000$, 
$c=0.5$, with $\varepsilon = 1/3$ and 
$\sigma$ = 2.1 (filled circles).
(b) Logarithmic scaling of $\tau_1$ versus $L$ 
(shown in log-lin scale)
for $\sigma = 2.5$ (filled circles). In this case $\sigma_\Lambda=1.99$
and $\sigma_V = 2.15$.
The values of $\tau_1$ have been computed
with $\Delta = 10^{-8}$ and averaged over
$10^3$ initial conditions.}
\label{f5}
\end{figure}

\begin{figure}[h]
\cl{
\psfig{figure=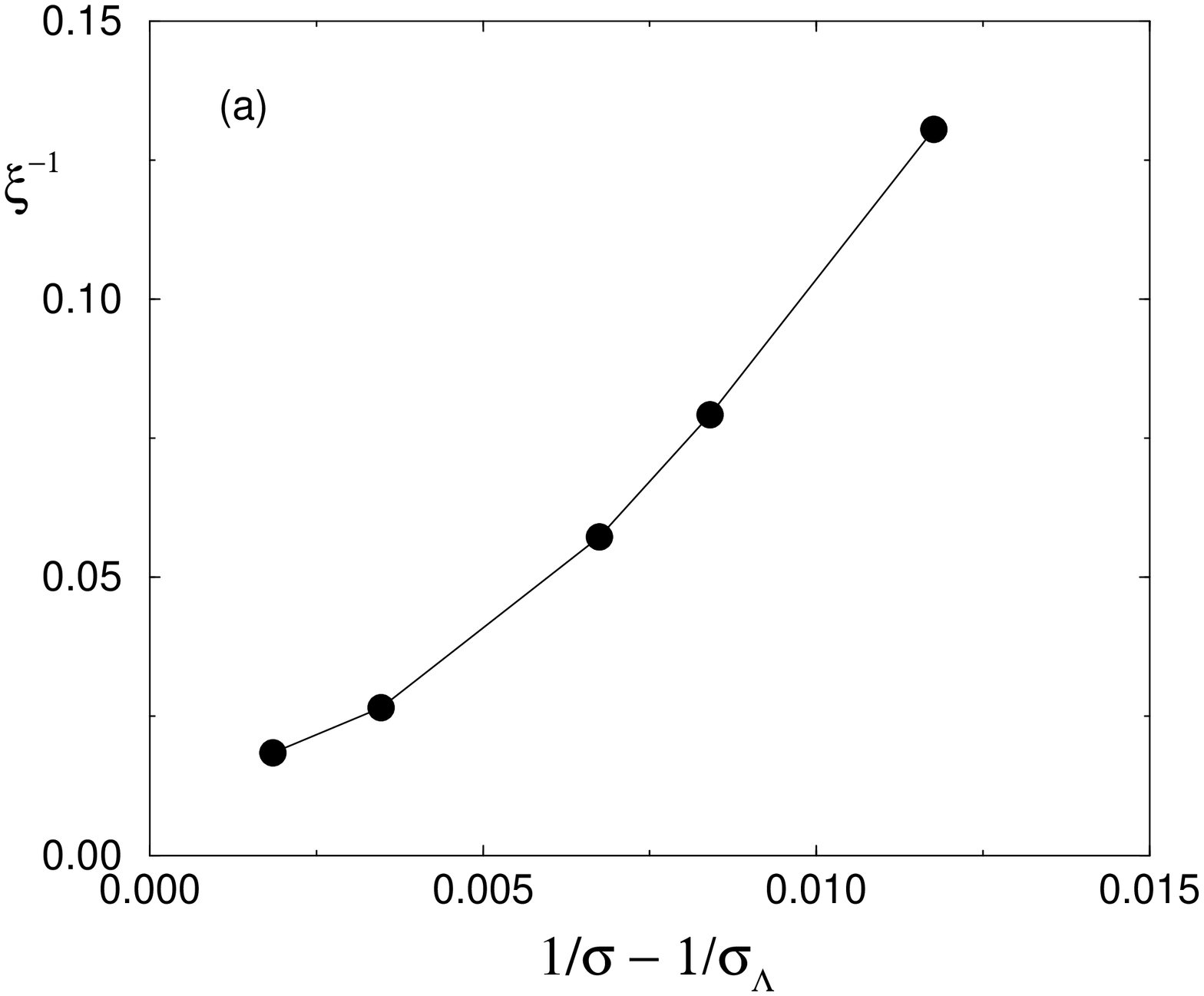,angle=-90,width=7truecm,height=7truecm}
\psfig{figure=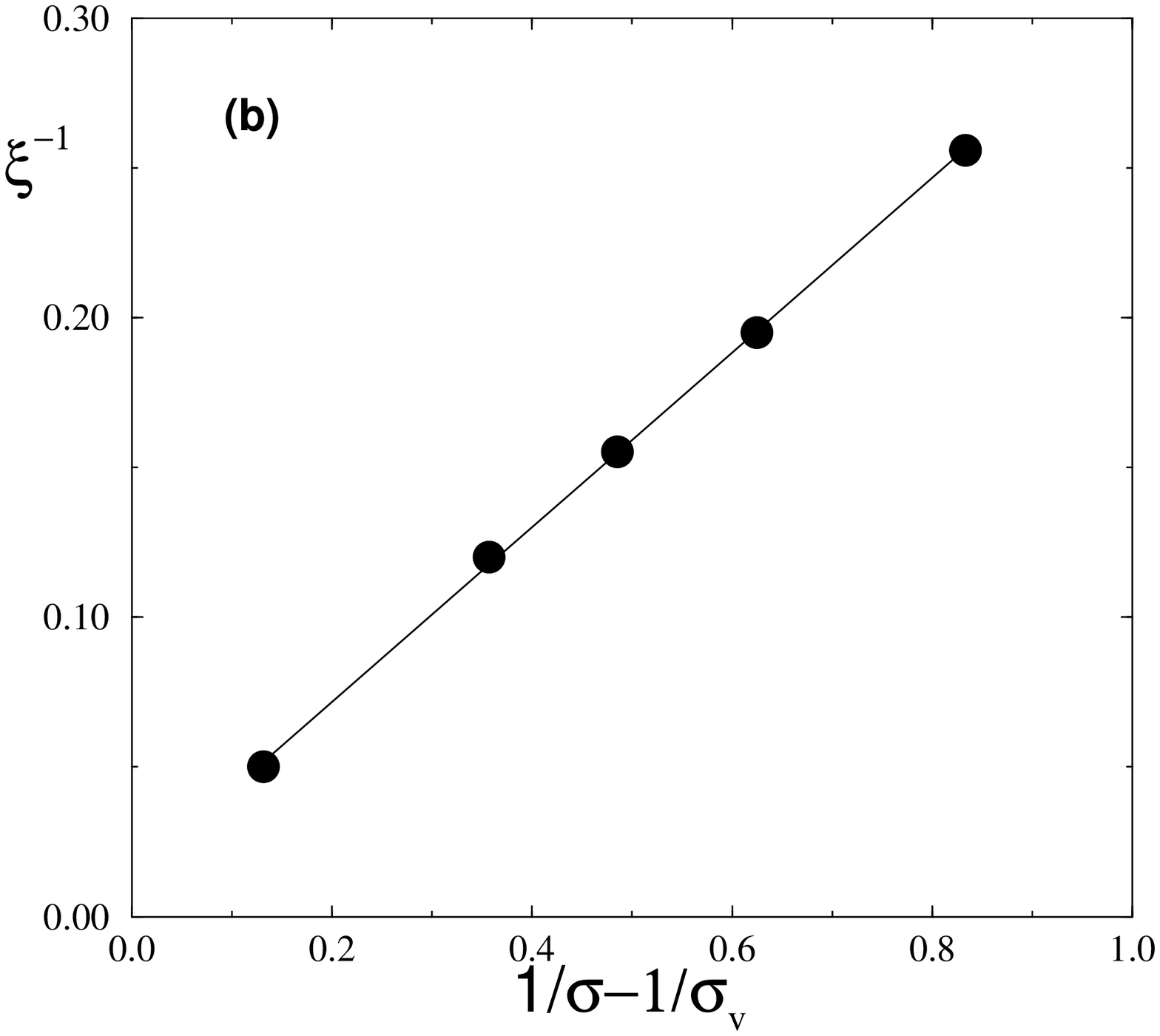,angle=-90,width=7truecm,height=7truecm}
}
\vskip 0.7 cm
\caption{
(a) The inverse of the scaling factor $1/\xi$ as a function of
$\sigma^{-1} - \sigma_\Lambda^{-1}$ ($\sigma_\Lambda = 2.5015$) 
is reported for coupled
(\protect\ref{contmap}) maps with $b=2.0$ and
$\varepsilon = 1/3$ and $2.49 \le \sigma \le 2.43$;
(b) $1/\xi$ as a function of
$\sigma^{-1} - \sigma_V^{-1}$ ($\sigma_V = 0.4018$) 
for coupled logistic maps with
$\varepsilon = 2/3$ and $0.30 \le \sigma \le 0.38$.}
\label{f3}
\end{figure}

\begin{figure}[h]
\cl{
\psfig{figure=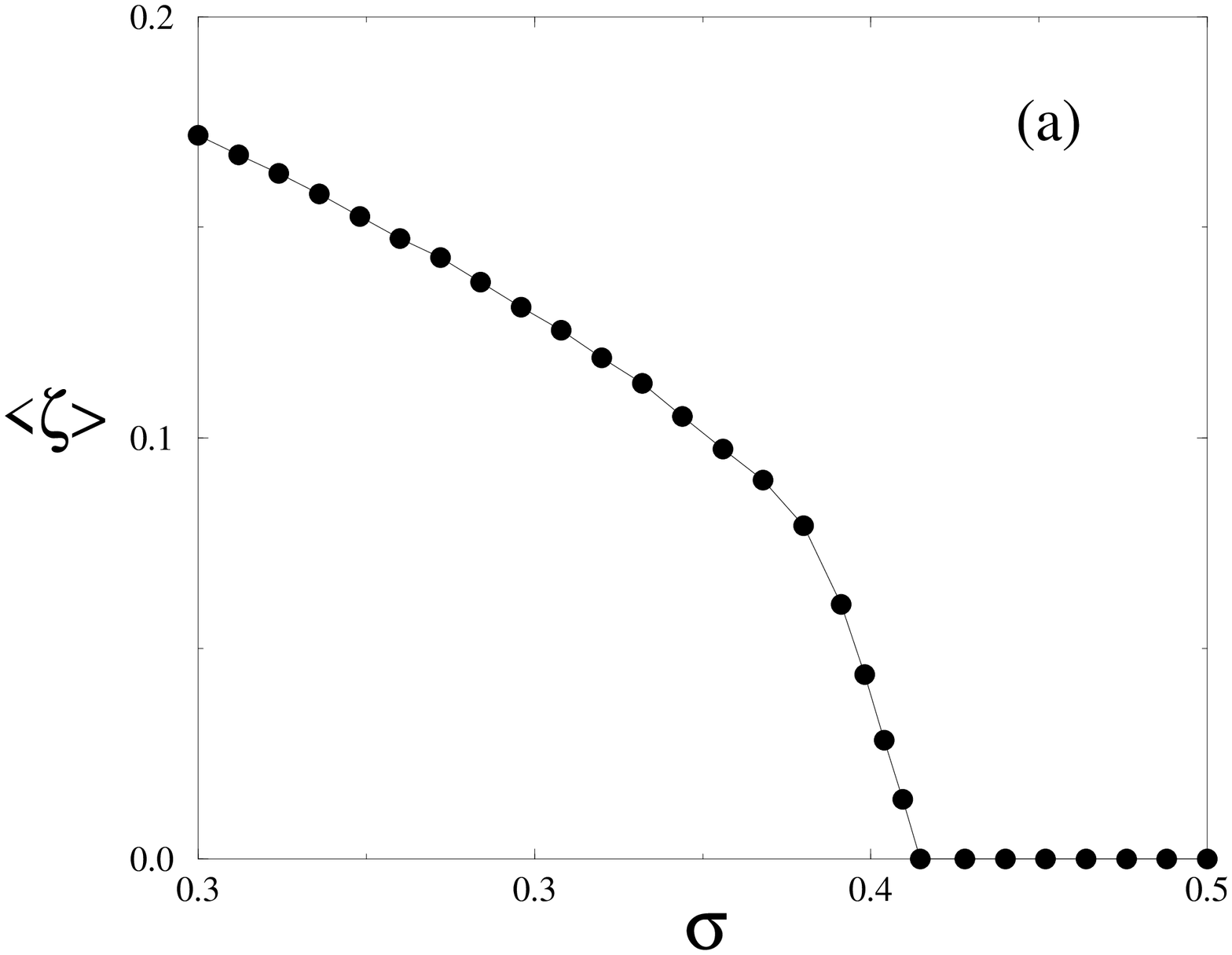,angle=-90,width=7truecm,height=7truecm}
\psfig{figure=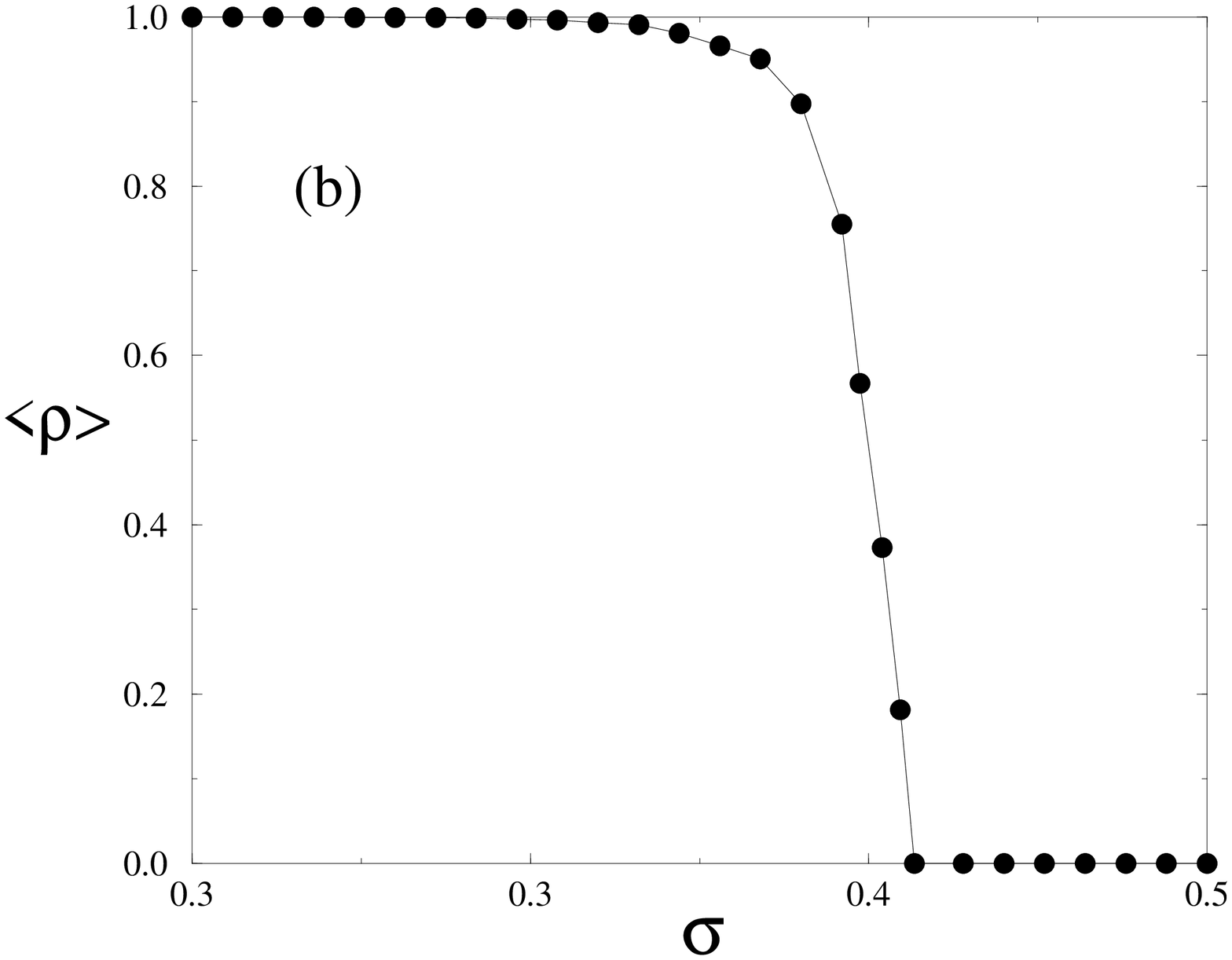,angle=-90,width=7truecm,height=7truecm}
}
\vskip 0.7 cm
\caption{The spatio-temporal averages of the 
indicators $\zeta$ (a) and $\rho$ (b) are reported as a function of
the noise amplitude $\sigma$ for for coupled logistic maps with
$\varepsilon = 2/3$.
The data have been
obtained by averaging over a time $t=10^4$ and 20 different
initial conditions, for $L=4096$ and considering a threshold 
$\Delta=10^{-12}$.
}
\label{f8}
\end{figure}

\begin{figure}[h]
\cl{
\psfig{figure=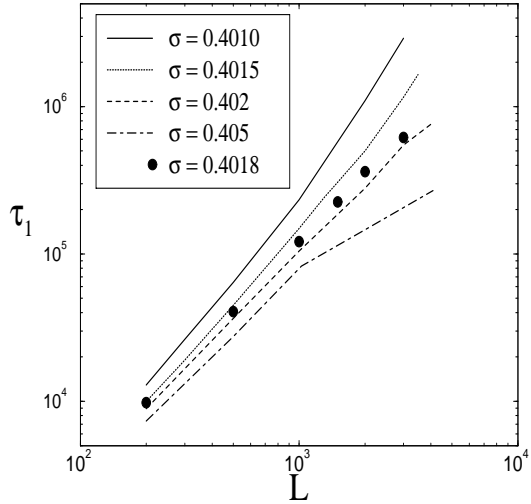,angle=-90,width=7truecm,height=7truecm}
}
\vskip 0.7 cm
\caption{The first passage time $\tau_1$ is reported in a log-log scale
as a function of system size $L$ for coupled logistic maps with
$\varepsilon = 2/3$ and for various $\sigma$. The data have been
obtained by averaging over $3,000$ - $25,000$ different initial conditions
and considering a threshold $\Delta=10^{-12}$.
}
\label{f9}
\end{figure}

\begin{figure}[h]
\cl{
\psfig{figure=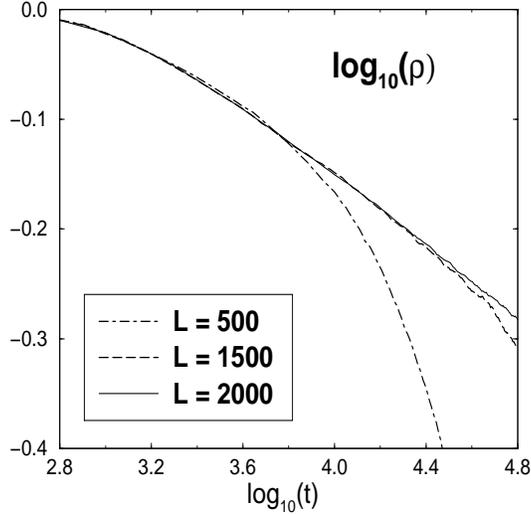,angle=-90,width=7truecm,height=7truecm}
}
\vskip 0.7 cm
\caption{Logarithm in base 10 of the density of active sites as a function of 
$\log_{10} (t)$ for coupled logistic maps with
$\varepsilon = 2/3$ and $\sigma = 0.4018$. The data have been
obtained by averaging over $10,000$ different initial conditions
and considering a threshold $\Delta=10^{-15}$.
}
\label{f10}
\end{figure}

\begin{figure}[h]
\cl{
\psfig{figure=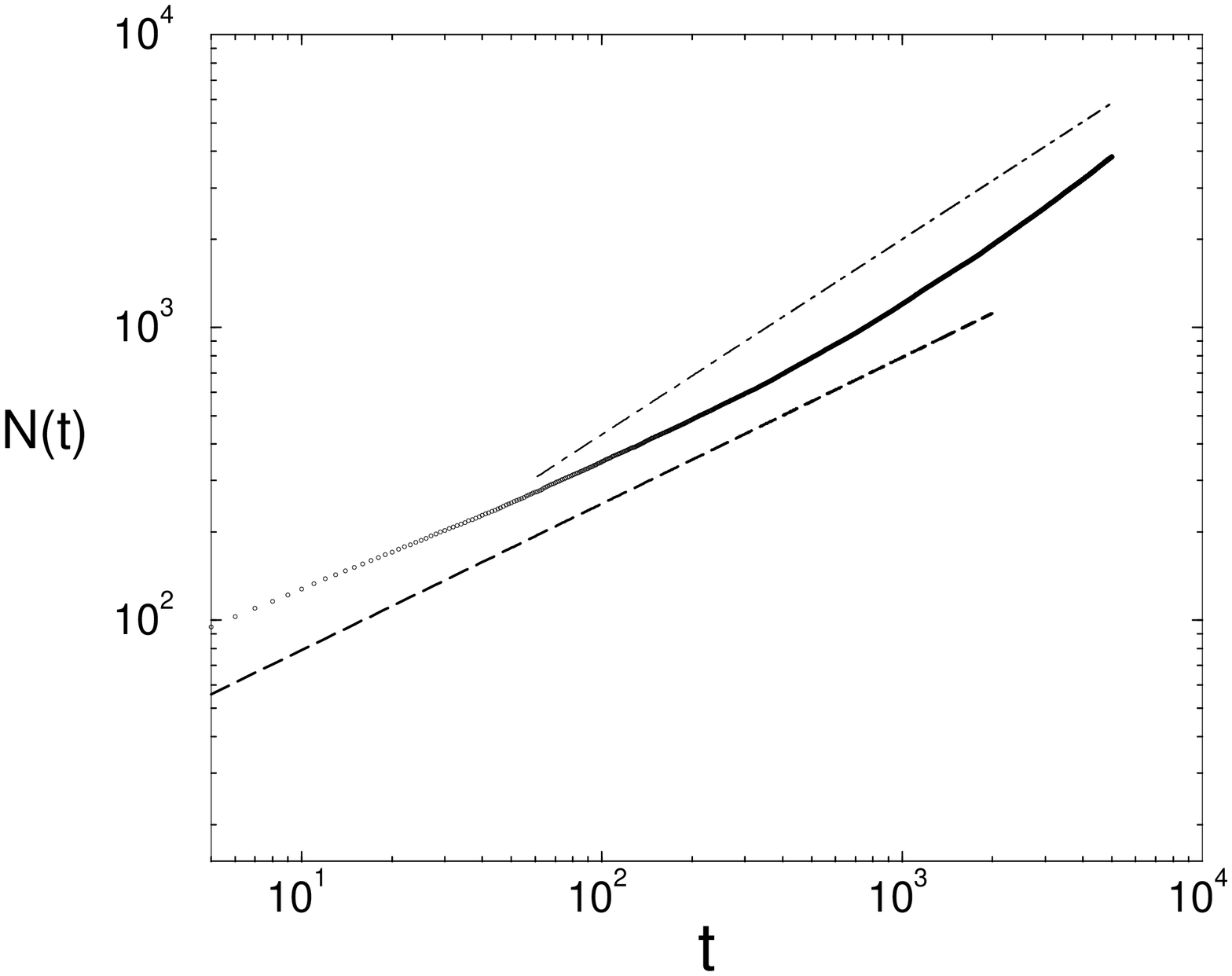,angle=-90,width=7truecm,height=7truecm}
}
\vskip 0.7 cm
\caption{The number of ``infected'' sites $N(t)$ is 
reported in a log-log scale
as a function of time $t$ for coupled maps (\protect\ref{contmap}) with
$a=4$,$b=2$,$c=0.5$ and $\varepsilon = 1/3$ and for various $\sigma$. 
The numerical data (filled circles)
have been obtained for a chain length $L=20,000$
by averaging over $1,000$ different initial conditions
and considering a threshold $\Delta=10^{-15}$.
The dashed line indicates a scaling of the
Edwards-Wilkinson type (with exponent $1/z=0.5$)
while the dot-dashed line has a slope $1/z=2/3$
analogous to that of the KPZ scaling.
}
\label{f11}
\end{figure}


\begin{thebibliography}{10}
\bibitem{nota1}  We have checked that different norms, e.g.
the maximum or the euclidean ones, do not modify the scenario
described herein.
Note that both quantities are well defined also 
in the noise-free case, since there is a finite 
probability that two different initial conditions may stay closer than 
a finite distance $\Delta$.
\end{thebibliography}
\end{document}